\documentclass[preprint,review,12pt]{elsarticle}

\usepackage{amssymb}

\def\be{\begin{equation}}
\def\ee{\end{equation}}

\def\bi{\begin{itemize}}
\def\ei{\end{itemize}}
\def\bn{\begin{enumerate}}
\def\en{\end{enumerate}}
\def\bea{\begin{eqnarray}}
\def\eea{\end{eqnarray}}
\def\no{\nonumber}
\def\ba{\begin{array}}
\def\ea{\end{array}}
\def\bd{\begin{displaymath}}
\def\ed{\end{displaymath}}

\journal{Physics Letters A}

\begin{document}

\begin{frontmatter}



\title{Quantum renormalization group approach to geometric phases in spin chains}


\author{R. Jafari}

\address{Research Department, Nanosolar System Company (NSS), Zanjan 45158-65911, Iran}
\address{Department of Physics, Institute for Advanced
Studies in Basic Sciences (IASBS), Zanjan 45137-66731, Iran}


\begin{abstract}
A relation between geometric phases and criticality of spin chains
are studied using the quantum renormalization-group approach.
I have shown how the geometric phase evolve as the size of
the system becomes large, i.e., the finite size scaling is
obtained. The renormalization scheme demonstrates how the first
derivative of the geometric phase with respect to the field strength
diverges at the critical point and maximum value of the
first derivative, and its position, scales with the exponent of the system size.
\end{abstract}


\end{frontmatter}

{Email address: jafari@iasbs.ac.ir, rohollah.jafari@gmail.com\\}
{Tel: (+98) 241-4152118, Fax: (+98) 241 4244949}

\section{Introduction}
\label{Introduction}

Quantum phase transition (QPT) has been one of the most interesting topics
in the area of strongly correlated systems \cite{Vojta}.
At zero temperature, the properties of the ground state
may change drastically showing a non-analytic behavior of a physical
quantity by reaching the quantum critical point. This can be done by
tuning a parameter in the Hamiltonian, for instance, the magnetic
field or the amount of disorder.	
Traditionally such a problem is addressed by resorting to
notions such as order-parameter and symmetry breaking i.e.,
the Landau-Ginzburg paradigm \cite{Goldenfeld}. In the last few years
a big effort has been devoted to the analysis of QPTs from
the Quantum Information perspective \cite{Osterloh,kargarian1,kargarian2,Jafari1,kargarian3,Jafari2,Langari},
the main tool being the study of different entanglement measures
\cite{Osborne}.
In the view of some difficulties \cite{Reuter}, attention has shifted to include
other, potentially related, means of characterizing QPTs \cite{Zanardi}.
One such approach centers around the notion of geometric phase (GP).
GP has been offered as a typical mechanism for a quantum system to keep
the memory of its evolution in Hilbert space. Such phases were introduced
in quantum mechanics by Berry in 1984 \cite{Berry}. Since then,
geometric phases became objects of theoretical and experimental
researches \cite{Shapere} uncovering that they are related to a number
of important physical phenomena \cite{Bohm} such as Aharonov-Bohm \cite{Aharonov}
and quantum Hall effects \cite{Klitzing}.
In recent years, this interest is increased due to their applicability
in quantum-information processing \cite{Zanardi2}.
In other words, GP has become extendable to product states
of composite systems since the uncorrelated subsystems pick up independent
geometric phase factors. However, GP could be induced by quantum entanglement,
if the full state is pure. On the other hand, classical correlations and quantum
entanglement can coexist in mixed quantum states, which means the
forms of the mixed state of the geometric phases \cite{Uhlmann}, applied to the path of the relative
states, may contain portions from both types of correlations.
Nevertheless, their connection to the quantum phase transitions has been manifested
recently in Ref.\cite{Carollo,Zhu,Hamma}, where it is shown that the geometric phase could be used to
investigate the critical properties of the spin chains \cite{Carollo}. On the other hand, the critical
exponents can be evaluated from the scaling behavior of the geometric phases
\cite{Zhu}. Therefore, the geometric phase could be considered as a topological test
for manifestation of quantum phase transitions \cite{Hamma}.
These general relation originates from topological property of the geometric phase. It describes the curvature of the Hilbert space and is directly
related to the degeneracy property in the quantum systems.
The degeneracy in the many-body systems plays crucial role
in our understanding of the quantum phase transition.
Thus the geometric phase can be considered as another powerful tool for
detecting the QPT.

Our main purpose in this work is to hire quantum
renormalization group (QRG) \cite{Pfeuty1} to study the evolution of
the geometric phase of spin models. To have a concrete discussion,
the one dimensional $S=\frac{1}{2}$ Ising model in transverse
field (ITF) is considered by implementing the quantum
renormalization group approach \cite{kargarian1,kargarian2,Jafari1,Jafari2,kargarian3,Langari,miguel1,Jafari3,miguel2}.
To the best of my knowledge, the GP properties study has only been done for exactly solvable models and this is the first report which addresses how to get GP properties
of the models which are not exactly solvable using QRG. I also show that QRG-based investigation of
the GP of the models is more convenient and also accurate than that of entanglement (concurrence).


\section{Theoretical Model}
\label{TM}

Consider the ITF model
on a periodic chain of $N$ sites with Hamiltonian
\be
\label{eq1}
H=-J\sum_{i=1}^{N}(\sigma_{i}^{x}\sigma_{i+1}^{x}+\lambda\sigma_{i}^{z}),
\ee
where $J>0$ and $\lambda$ are the exchange coupling and the transverse
field, respectively. From the exact solution \cite{miguel2,Pfeuty2} it can be seen that a
second order phase transition occurs for $\lambda_{c}=1$ where
the behavior of the order parameter or magnetization is given
by $<\sigma^{x}>=(1-\lambda)^{1/8}$ for $\lambda<1$ and $<\sigma^{x}>=0$ for
$\lambda>1$.

\section{Quantum Renormalization Group}
\label{qrg}

The main idea of the RG method is the mode elimination or thinning
of the degrees of freedom followed by an iteration which reduces
the number of variables step by step till reaching a fixed point.
In Kadanoff's approach, the first step of the QRG method consists
of assembling a set of lattice points into disconnected blocks of
$n_{B}$ sites.
In this fashion, the total number of blocks in the whole chain would be $N'=N/n_{B}$. This partitioning of the lattice into blocks induces
a decompositioning of the Hamiltonian into two parts: intra-block ($H_{B}$) and inter-block ($H_{BB}$) Hamiltonians. The block Hamiltonian
${H_B}$ is a sum of commuting Hamiltonians ($h_{B}$) acting on individual blocks. The diagonalization of $h_{B}$ for small
$n_{B}$ is achieved analytically and then intra-block Hamiltonian and inter-block Hamiltonian
is projected into the low energy subspace of $H_{B}$. Afterwards, the original Hamiltonian
is mapped into an effective Hamiltonian ($H^{eff}$) which acts on
the renormalized subspace \cite{miguel1,Jafari3,miguel2}.

In this paper, to implement QRG, the Hamiltonian is divided into two-site blocks

\bea
\no
H^{B}=\sum_{I=1}^{N/2}h_{I}^{B},~h_{I}^{B}=-J(\sigma_{1,I}^{x}\sigma_{2,I}^{x}+\lambda\sigma_{1,I}^{z}),
\eea
and the remaining part of the Hamiltonian is included in the
inter-block part

\bea
\no
H^{BB}=-J\sum_{I=1}^{N/2}(\sigma_{2,I}^{x}\sigma_{1,I+1}^{x}+\lambda\sigma_{2,I}^{z}),
\eea
where $\sigma_{j,I}^{\alpha}$ refers to the $\alpha$-component of
the Pauli matrix at site $j$ of the block labeled by $I$. The
Hamiltonian of each block ($h_{I}^{B}$) is diagonalized exactly
and the projection operator

\bea
\label{eq2}
P_{0}=|\psi_{0}\rangle\langle\psi_{0}|+|\psi_{1}\rangle\langle\psi_{1}|,
\eea
is constructed from the two lowest eigenstates in which $|\psi_{0}\rangle$ is the ground state and $|\psi_{1}\rangle$ is
the first excited state. In this respect the effective Hamiltonian

\bea
\no
H^{eff}=P_{0}[H^{B}+H^{BB}]P_{0}
\eea
is matched to the original one (Eq.(\ref{eq1})) replacing the couplings
with the following renormalized coupling constants.
\bea \label{eq3}
J'=J\frac{2q}{1+q^{2}},~q=\lambda+\sqrt{\lambda^{2}+1},
~\lambda'=\lambda^{2}.
\eea

\begin{figure}[h]
\begin{center}
\includegraphics[width=8.5cm]{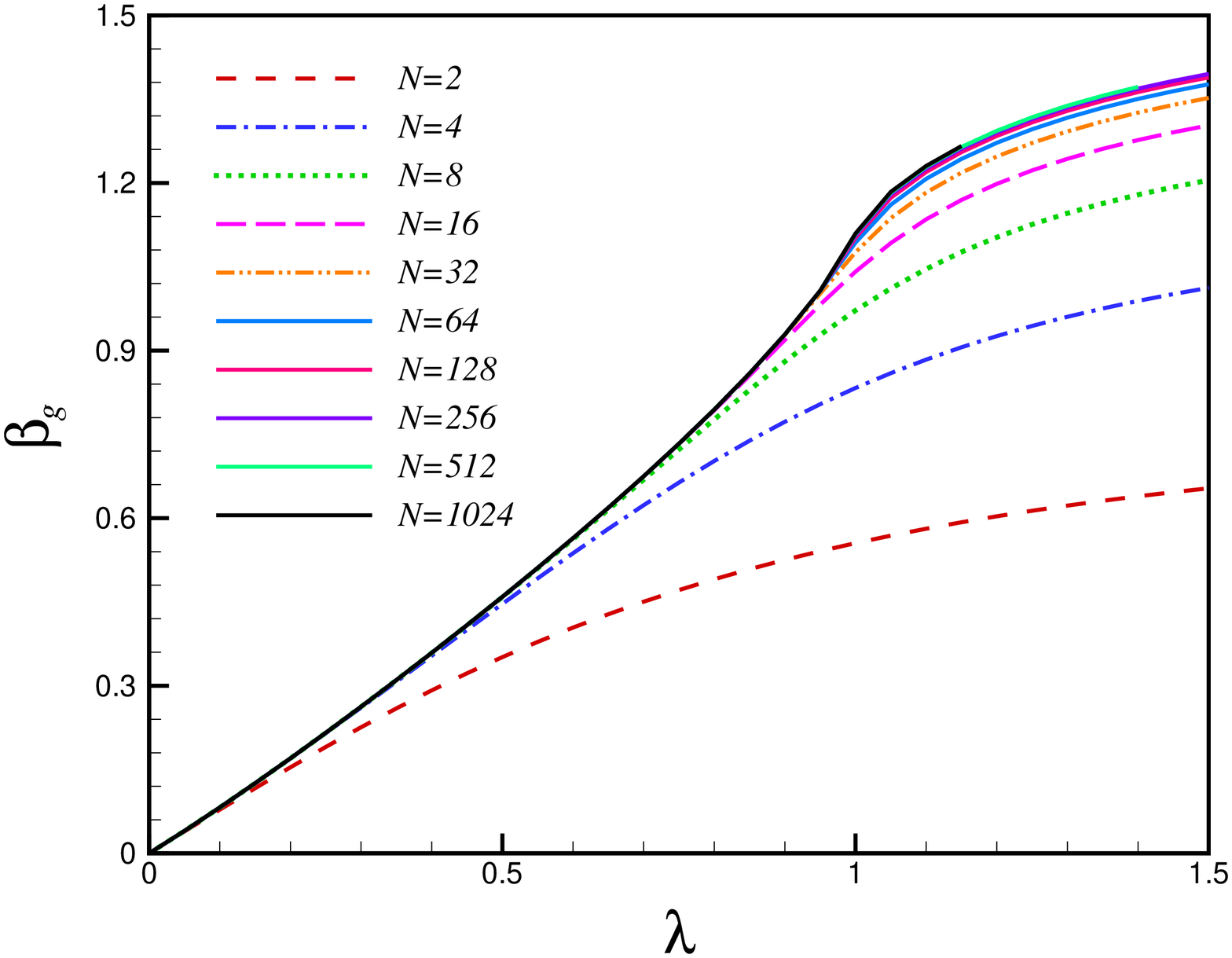}
\caption{(Color online) Evolution of the geometric phase under RG
versus $\lambda$.} \label{fig1}
\end{center}
\end{figure}


\section{Geometric Phase and Renormalization Group Application}
\label{GP}

To investigate the geometric phase in systems, a new
family of Hamiltonians are introduced that can be described by applying a
rotation of $\phi$ around the $z$ direction to each spin \cite{Carollo}, i.e.,

\bea
\label{eq4}
H_\phi=g_\phi^\dagger H g_\phi;~~~ g_\phi=\prod_{j=1}^N
\exp(-i\phi\sigma_l^z/2).
\eea
The critical behavior is independent of $\phi$ as the spectrum of the system is $\phi$
independent \cite{Zhu}.
The geometric phase of the ground state, accumulated by varying the
angle $\phi$ from $0$ to $\pi$, is described by

\bea
\label{eq5}
\beta_g=-i\int_0^\pi\langle\psi_{\phi}|\frac{\partial}{\partial\phi}|\psi_{\phi}\rangle d\phi,
\eea
here $|\psi_{\phi}\rangle$ is the ground state of $H_\phi$ \cite{Carollo}.

\begin{figure}[h]
\begin{center}
\includegraphics[width=8.5cm]{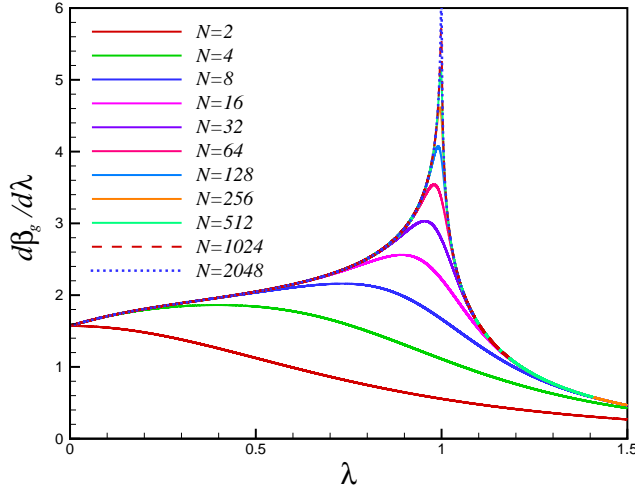}
\caption{(Color online) The first derivative of geometric
phase for different system size. For the limit of large system
(high RG step), the non-analytic behavior of the first
derivative of GP is obtained through the diverging.} \label{fig2}
\end{center}
\end{figure}

The eigenvalues of the Hamiltonian $H$ will not affected
by this unitary transformation. So the eigenvectors of
new Hamiltonians $H_\phi$ can be obtained by acting the rotation operator on
the eigenvectors of the former Hamiltonian ($H$).
In other words, $|\psi_{\phi}\rangle=g_\phi|\psi\rangle$ where
$|\psi\rangle$ and $|\psi_{\phi}\rangle$ are the eigenvectors of $H$ and $H_\phi$,
respectively. However, the projection operators of new Hamiltonian
$H_{\phi}$(($P_{0}(\phi)$) and the unrotated Hamiltonian (Eq. (\ref{eq2}))
are related by

\bea
\no
P_{0}(\phi)=g_\phi^\dagger P_{0} g_\phi.
\eea

On the other hand, the ground state of the renormalized chain $|\psi'\rangle$
will be related to that of the original one by the
transformation $|\psi\rangle=P_{0}|\psi'\rangle$.
It is straightforward to show that the geometric phase
in the renormalized chain is described by

\bea
\label{eq6}
\beta_{g}^{n+1}=\beta_{g}^{n}-\frac{\pi}{2}\frac{\gamma^{n}}{2}
\eea

where $\beta_{g}^{n}$ is geometric phase at the $nth$ step of RG
and $\gamma^{0}$ is defined by $\frac{q^{2}-1}{q^{2}+1}$.
The expression for $\gamma^{n}$ is similar to $\gamma^{0}$ where the
coupling constants should be replaced by the renormalized
ones at the corresponding RG iteration ($n$).
In this approach, geometric phase at each iteration of RG is connected to
its value after a RG iteration by Eq. (\ref{eq6}).
This will be continued till reaching a controllable fixed point
where the value of the geometric phase could be obtained \cite{Jafari1,Jafari3}.

\section{Numerical Results: Scaling properties of Geometric Phase}
\label{NR}

\begin{figure}[t]
\begin{center}
\includegraphics[width=8.5cm]{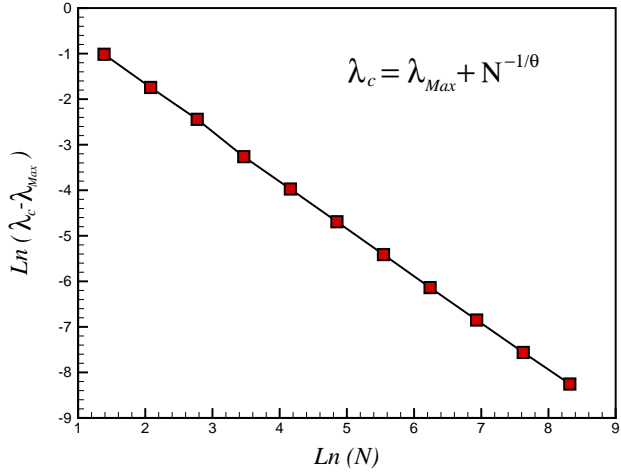}
\caption{(Color online) Scaling  of the position ($\lambda_{Max}$) of
$\frac{d\beta_{g}}{d\lambda}$ for different length chains.
$\lambda_{Max}$ goes to $\lambda_{c}$ as
the size of the system increase as $\lambda_{c}=\lambda_{Max}+N^{-1/0.957}$.}
\label{fig3}
\end{center}
\end{figure}

In this section the numerical results of the model would be discussed.
The evolution of $\beta_g$ under RG steps versus $\lambda$ is presented
in Fig. \ref{fig1}. In the $nth$ RG step the expression given in Eq. (\ref{eq6}) is
evaluated at the renormalized coupling given by the $n$ iteration
of $\lambda$ given in Eq. (\ref{eq3}). The zero RG step means a bare
two-site model, while in the first RG step the effective two-site
model represents a four-site chain.
Generally, in the $nth$ step of RG, a chain of $2^{n+1}$ sites is represented effectively by
the two sites with renormalized couplings. All curves in Fig. (\ref{fig1})
have a kink at the critical point, $\lambda_{c}=1$, for large systems.
At the critical point, correlation length is infinite and fluctuations
occur on all length scales which means that the system is scale-invariant.
The non-analytic behavior is a feature of second-order quantum phase transition.
It is also accompanied by a scaling behavior since the correlation length diverges and
there is no characteristic length in the system at the critical
point.

\begin{figure}[h]
\begin{center}
\includegraphics[width=8.5cm]{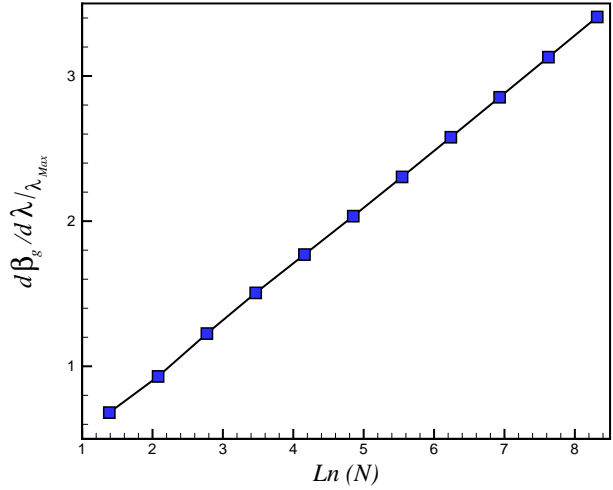}
\caption{(Color online) Scaling the maximum of
$\frac{d\beta_{g}}{d\lambda}$ for various size of system, and
the maximum diverges as
$\frac{d\beta_{g}}{d\lambda}|_{\lambda_{Max}}\approx 0.393\ln N$.}
\label{fig4}
\end{center}
\end{figure}

\begin{figure}[h]
\begin{center}
\includegraphics[width=8.5cm]{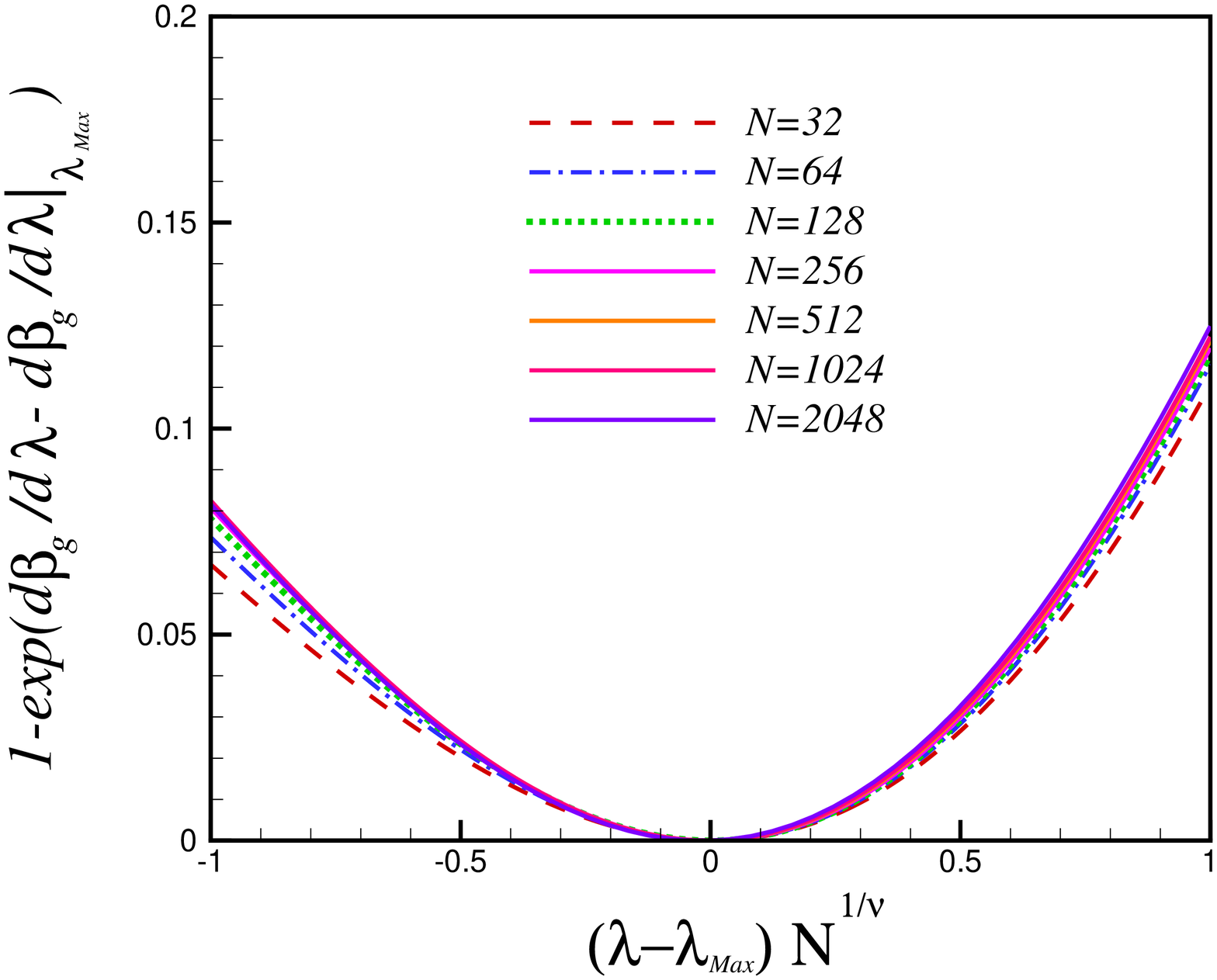}
\caption{(Color online) Finite-size scaling for different
lattice sizes through the RG treatment. The curves which correspond
to different system sizes clearly collapse on a single curve.}
\label{fig5}
\end{center}
\end{figure}

Zhu has verified that the GP of ground state in the XY model
in the transverse field obeys scaling behavior in the vicinity of a
quantum phase transition \cite{Zhu}.
In particular he has shown that the geometric phase is non-analytical and
its derivative with respect to the magnetic field diverges at the critical point.
As it is previously stated, a large system, i.e. $N=2^{n+1}$, can
be effectively describe by two sites with the renormalized coupling in the $nth$ RG step.
The first derivative of GP is analyzed as a function of magnetic
field at different RG steps which manifest the size of the system.
In Fig. (\ref{fig2}) the derivative of GP with respect to the coupling
constant $d\beta_{g}/d\lambda$ is presented which, shows a singular behavior at the critical
point as the size of system becomes large. This singular behavior is the result of
the kink in GP at $\lambda=\lambda_{c}$ (Fig. (\ref{fig1})).

It is found that the position of the maximum
of  $d\beta_{g}/d\lambda$ ($\lambda_{Max}$) tends towards the critical
point like

\bea
\no
\lambda_{c}=\lambda_{Max}+N^{-1/\theta},
\eea
where $\theta\simeq0.957$ (Fig. (\ref{fig3})).
Moreover, the scaling
behavior of $\frac{d\beta_{g}}{d\lambda}|_{\lambda_{Max}}$ versus $\ln N$ is derived. This
quantity is shown in Fig.(\ref{fig4}), which behaves linearly and
the scaling behavior is obtained as

\bea
\no
\frac{d\beta_{g}}{d\lambda}|_{\lambda_{Max}}\approx\kappa\ln N
\eea
with $\kappa=0.393$.

The exponent $\theta$ is directly related
to the correlation length exponent ($\nu$) close to the critical
point. The correlation length exponent gives the behavior of
the correlation length in the vicinity of $\lambda_{c}$, i.e., $\xi\sim
(\lambda-\lambda_{c})^{-\nu}$. Under the RG transformation of Eq. (\ref{eq3}),
the correlation length scales in the $n$th RG step as
$\xi^{(n)}\sim (\lambda_{n}-\lambda_{c})^{-\nu}=\xi/n_{B}^{n}$, which
immediately leads to an expression for $|\frac{d\lambda_{n}}{d\lambda}|_
{\lambda_{c}}$ in terms of $\nu$ and $n_{B}$. Dividing the last equation
into $\xi\sim (\lambda-\lambda_{c})^{-\nu}$
gives rise to $|\frac{d\lambda_{n}}{d\lambda}|_{\lambda_{c}}\sim N^{1/\nu}$ , which implies that
$\theta=1/\nu$, since $\frac{d\beta_{g}}{d\lambda}|_{\lambda_{Max}}\sim
|\frac{d\lambda_{n}}{d\lambda}|_{\lambda_{c}}$ at the critical point. We should
note that the scaling of in the the position of $\lambda_{Max}$
(Fig. (\ref{fig3})), comes from the divergence of the
correlation length near the critical point.
In the large system size limit, when approaching to the critical point the
correlation length almost covers the size of the system
$\xi\sim N$ which results in the following scaling form

\bea
\no
\lambda_{c}=\lambda_{Max}+N^{-1/\nu}.
\eea

To obtain the finite-size scaling behavior of
$\frac{d\beta_{g}}{d\lambda}|_{\lambda_{Max}}$, we look for a scaling function
when all graphs tend to collapse on each other under
RG evolution which results in a large system. This is also a
manifestation of the existences of the finite size scaling for the
GP. Fig. (\ref{fig5}) shows the plot of $1-exp\big(\frac{d\beta_{g}}{d\lambda}-\frac{d\beta_{g}}{d\lambda}|_{\lambda_{Max}}\big)$ versus
$N(\lambda-\lambda_{Max})$. The lower curves, which are for large
system sizes, clearly show that all plots fall on each other.

The similar scaling behaviors as well as their relation to correlation
length exponent have been reported in our previous works
\cite{kargarian1,Jafari2}, where the static properties of
the ground state entanglement and low energy state dynamics of entanglement
of ITF model by RG method were studided. These facts strongly imply the important
relation between quantum entanglement and geometric phase, and provides a
possible understanding of entanglement from the topological structure of the
systems. This point can be understood by noting that both of the mentioned methods are connected to
the correlation functions, and also are connected directly to each other by the
inequality \cite{Cui}.

\section{Summary}
\label{S}
To summarize, the idea of renormalization group
(RG) to study the geometric phase of Ising model in transverse field is implemented.
In order to explore the critical behavior of the ITF model the evolution
of geometric phase through the renormalization of the lattice were examined.
In this respect I have shown that the RG procedure can be implemented
to obtain the GP of a system and its finite size scaling in terms
of the effective Hamiltonian which is described by the
renormalized coupling constants.
The phase transition becomes significant which
shows a diverging behavior in the first derivative of the geometric phase.
This divergence of GP are accompanied by a scaling behavior near the critical point
where the size of the system becomes large. The scaling behavior
characterizes how the critical point of the model is
touched as the system size is increased. It is also shown that the non-analytic
behavior of GP is originated from the correlation length exponent in the
vicinity of the critical point. This shows that the behavior of the GP
near the critical point is directly connected to the quantum critical
properties of the model. We get the properties of GP for a large
system dealing with a small block which make it possible to
get analytic results.
However, the numerical results of QRG show that the application of QRG
to manifest the GP properties, is quantitatively more accurate than its application
on quantum information resource \cite{kargarian1,kargarian2,Jafari1,Jafari2,kargarian3}.
\section{acknowledgments}
The author would like to thank S. N. S. Reihani, A. Akbari and A. Langari for reading the
manuscript, fruitful discussions and comments.





\bibliographystyle{model1a-num-names}
\bibliography{<your-bib-database>}

\begin{thebibliography}{00}


\bibitem{Vojta}
M. Vojta, Rep. Prog. Phys. \textbf{66}, (2003) 2069 and references therein.

\bibitem{Goldenfeld}
N. Goldenfeld, \emph{Lectures on phase transitiosn and the
renormalization group}, Westview Press, Boulder, 1992.

\bibitem{Osterloh}
A. Osterloh, Luigi Amico, G. Falci and Rosario Fazio, Nature
\textbf{416}, 608 (2002).

\bibitem{kargarian1}
M. Kargarian, R. Jafari and A. Langari, Phys. Rev. A \textbf{76},
060304(R)(2007).

\bibitem{kargarian2}
M. Kargarian, R. Jafari and A. Langari, Phys. Rev. A \textbf{77},
032346 (2008).

\bibitem{Jafari1}
R. Jafari, M. Kargarian, A. Langari, and M. Siahatgar, Phys. Rev.
B \textbf{78}, 214414 (2008).

\bibitem{kargarian3}
M. Kargarian, R. Jafari and A. Langari, Phys. Rev. A \textbf{79},
042319 (2009).

\bibitem{Jafari2}
R. Jafari, Phys. Rev. A \textbf{82}, 052317 (2010).

\bibitem{Langari}
A. Langari, A. T. Rezakhani, New J. Phys. \textbf{14}, 053014 (2012);
N. Amiri, A. Langari,  physica status solidi (b). \textbf{250}, 417 (2013).


\bibitem{Osborne}
T.J. Osborne, M.A. Nielsen,  Phys. Rev. A. \textbf{66}, 032110
(2002); G. Vidal, J.I. Latorre, E. Rico, and A. Kitaev,
Phys. Rev. Lett. \textbf{90}, 227902 (2003); Y. Chen,
P. Zanardi, Z. D. Wang, F. C. Zhang, New J. Phys. 8,
\textbf{97} (2006); L.-A. Wu, M.S. Sarandy, D.
A. Lidar, Phys. Rev. Lett. \textbf{93}, 250404 (2004)

\bibitem{Reuter}
M.E. Reuter, M.J. Hartmann, M.B. Plenio, Proc. Roy.
Soc. Lond. A \textbf{463}, 1271 (2007).


\bibitem{Zanardi}
P. Zanardi and N. Paunkovic, Phys. Rev. E \textbf{74}, 031123 (2006);
A. T. Rezakhani, P. Zanardi, Phys. Rev. A \textbf{73}, 012107 (2006);
\textbf{73}, 052117 (2006).


\bibitem{Berry}
M. V. Berry, Proc. R. Soc. London \textbf{392}, 45 (1984).

\bibitem{Shapere}
A. Shapere and F. Wilczek (Eds.), \textit{Geometric Phases in
Physics}, World Scientific, Singapore, 1989.


\bibitem{Bohm}
A. Bohm, A. Mostafazadeh, H. Koizumi, Q. Niu, and J.
Zwanziger, \textit{The Geometric Phase in Quantum Systems},
Springer, Berlin, 2003.


\bibitem{Aharonov}
Y. Aharonov and D. Bohm, Phys. Rev. \textbf{115}, 485 (1959).


\bibitem{Klitzing}
K. v. Klitzing, G. Dorda, and M. Pepper, Phys. Rev.
Lett. \textbf{45}, 494 (1980).


\bibitem{Zanardi2}
P. Zanardi and M. Rasetti, Phys. Lett. A \textbf{264}, 94 (1999);
J. A. Jones, V. Vedral, A. Ekert, and G. Castagnoli, Nature
(London) \textbf{403}, 869 (2000).



\bibitem{Uhlmann}
A. Uhlmann, Rep. Math. Phys. \textbf{24},  229 (1986);
E. Sjoqvist, A.K. Pati, A. Ekert, J.S. Anandan, M. Ericsson, D.K.L. Oi, V.
Vedral, Phys. Rev. Lett. \textbf{85} 2845 (2000).



\bibitem{Carollo}
A. Carollo, and J. K. Pachos, Phys. Rev. Lett. \textbf{95},
157203(2005).

\bibitem{Zhu}
S. L. Zhu, Phys. Rev. Lett. \textbf{96}, 077206(2006).

\bibitem{Hamma}
A. Hamma, e-print: quant-ph/0602091.

\bibitem{Pfeuty1}
P. Pfeuty, R. Jullian, K. L. Penson, in \textit{Real-Space
Renormalization}, edited by T. W. Burkhardt and J. M. J. van
Leeuwen (Springer, Berlin, 1982), Chap. 5.

\bibitem{miguel1}
M. A. Martin-Delgado and G. Sierra, Int. J. Mod. Phys. A
\textbf{11}, 3145 (1996).


\bibitem{Jafari3}
R. Jafari and A. Langari, Phys. Rev. B \textbf{76}, 014412 (2007);
Physica A \textbf{364}, 213 (2006); A. Langari, Phys. Rev. B
\textbf{69}, 100402(R) (2004).

\bibitem{miguel2}
M. A. Martin-Delgado and G. Sierra, Phys. Rev. Lett. \textbf{76},
1146 (1996).

\bibitem{Pfeuty2}
P. Pfeuty, ANNALS of Physics, \textbf{57}, 79 (1970).


\bibitem{Cui}
H.T. Cui, Y.F. Zhang, Eur. Phys. J. D, \textbf{51}, 393 (2009).











\end{thebibliography}

\end{document}